\documentclass[11pt,fleqn]{article}

\usepackage{amsfonts}
\usepackage{amsmath}
\usepackage{amssymb}
\usepackage{mathtools}
\usepackage{mathrsfs}
\usepackage{enumerate}
\usepackage{bbm}

\usepackage{pstricks}
\usepackage{subfig}
\usepackage{graphicx}
\usepackage{xcolor}
\usepackage{units}

\usepackage{caption}
\usepackage{booktabs}
\usepackage{textcomp}
\usepackage{array}
\usepackage{cite}
\usepackage{cancel}

\date{\today}

\newcommand{\del}{\partial}                            
\newcommand{\ol}[1]{\overline{#1}}                     
\newcommand{\klammer}[1]{\left(#1\right)}              
\newcommand{\real}{\Re\mathfrak{e}\hspace{1pt}}        
\newcommand{\e}{\textrm{e}}                            
\newcommand{\hs}{\hspace{1pt}}                        

\newcolumntype{z}[1]{>{\RaggedRight\hspace{0pt}}p{#1}}
\newcolumntype{w}[1]{>{\RaggedRight\hspace{0pt}}p{#1}}
\newcolumntype{v}[1]{>{\Centering\hspace{0pt}}p{#1}}

\DeclareCaptionLabelSeparator{mysep}{\hspace{3pt}:\hspace{3pt}}
\DeclareCaptionLabelFormat{mypiccap}{Fig.\hspace{3pt}{#2}}
\DeclareCaptionLabelFormat{mytabcap}{Tab.\hspace{3pt}{#2}}

\begin{document}

\date{}
\title{
\vskip 2cm
{\bf\huge Crosschecks for Unification at the LHC}\\[0.8cm]}

\author{{\sc\normalsize
Val\'eri L\"owen\footnote{E-mail: loewen@th.physik.uni-bonn.de} and Hans Peter Nilles\footnote{E-mail: nilles@th.physik.uni-bonn.de}\!
\!}\\[1cm]
{\normalsize Bethe Center for Theoretical Physics}\\
{\normalsize and}\\
{\normalsize Physikalisches Institut der Universit\"at Bonn}\\
{\normalsize Nussallee 12, 53115 Bonn, Germany}\\[1cm]
}
\maketitle \thispagestyle{empty}
\begin{abstract}
{
Experiments at the Large Hadron Collider (LHC) might test the picture of supersymmetric Grand Unification in particle 
physics. We argue that the identification of gaugino masses is the most promising step in this direction. Mass predictions
for gauginos are pretty robust and often related to the values of the gauge couplings constants. They might allow a meaningful crosscheck for grand unification, at least in simple schemes like gravity, anomaly or mirage mediation.
}
\end{abstract}

\clearpage

\tableofcontents
\newpage

\section{Introduction}

In the next few years experiments will explore the physics at the Tera scale and bring new insight on the standard model of particle physics as well as possible Ans\"atze of physics beyond the standard model. At the Tera scale the gauge group of the electro-weak interactions $\mathrm{SU(2)}\times\mathrm{U(1)}$ is broken and masses for $W^\pm$, $Z$ gauge bosons as well as quarks and leptons are generated. The Tera scale of 1 TeV is very small compared to another fundamental scale of physics: the Planck scale $M_{\rm P}\sim\unit[10^{19}]{GeV}$, and within the standard model one is concerned about the stability of that scale: the so-called hierarchy problem. This problem is at the origin of many attempts to consider physics beyond the standard model.

One popular attempt is based on supersymmetry (SUSY) with numerous new particles at the Tera scale. These new particles lead to a cancellation of the quadratic divergencies in the standard model and could thus explain the stability of the weak scale (at a TeV) compared to the Planck scale. An examination of the simplest model,
the minimal supersymmetric standard model (MSSM) leads to two serendipitous observations. The new particles at the Tera scale influence the evolution of coupling constants in a way exhibiting
\begin{itemize}
\item gauge coupling unification at a scale of $M_{\rm GUT}\sim\unit[10^{16}]{GeV}$
\item (partial gauge)-Yukawa coupling unification at $M_{\rm GUT}$.
\end{itemize}
These arguments are necessarily rather indirect, as they require the existence of a desert between Tera scale and GUT scale. Nonetheless, this observation appears so appealing that further studies and checks are of great interest.

So some of the early questions to be addressed in the analysis of LHC data
would be:
\begin{itemize}
\item do the new particles (if found) cancel quadratic divergencies of the standard model?
\item are they consistent with grand unification?
\end{itemize}
But even if these questions could be answered positively we still have to face the dilemma that the arguments for grand unification are indirect. It would be highly desirable to identify further independent arguments (from observation) for unification of gauge (and Yukawa) couplings. These are the questions we would like to
address in this paper. 

It is evident that such crosschecks for unification can only be found in simple schemes.
Models with many particles at intermediate scales will not allow clear conclusions and we have to hope that nature is kind enough to provide us with a simple (minimal) scheme. It is also obvious, that within these minimal schemes we have to concentrate on properties that are rather robust (i.\,e.\ more model dependent quantities should be
put aside in a first step). In addition we have to look for a situation where the ``predictions" from unification can be cast in a framework which is controlled by the parameters of the low energy effective theory (controllable unification).

We analyze these questions in a scheme based on supersymmetry. The simplest model would be the MSSM, where we have a distinct spectrum and we know the masses as well as the $\beta$- and $\gamma$-functions. Controllable unification would then describe properties of the model in terms of these quantities. The robust properties to
look for might be found in the spectrum of the supersymmetric partners, in particular the gaugino masses ($A$-terms and soft scalar masses seem to be more model dependent). The study of the soft terms requires knowledge about SUSY breakdown and its mediation to the MSSM. Again, we have to concentrate on simple schemes, with simple boundary conditions at the large scale.\footnote{Such controllable boundary conditions might also be desirable in order to avoid a potential flavor problem.} Thresholds at the GUT scale and/or intermediate scale might spoil the picture and make low energy crosschecks for unification impossible. In that sense we need some luck to be able to study low energy
controllable unification in a meaningful way. We need a simple model (e.\,g. the MSSM) and a simple basic scheme of SUSY breakdown with simple boundary conditions at the GUT scale. This eliminates many schemes. The simplest scheme that survives is gravity mediation with universal boundary conditions at the large scale. The most robust predictions for soft terms are gaugino masses (followed by $A$-terms and soft scalar masses in that order).

So the main focus should be gaugino masses. In a scheme with universal gaugino masses $m_{1/2}$ at the GUT scale we still have to consider two cases: $m_{1/2} \neq 0$ and $m_{1/2} = 0$. The first case will correspond to ``gravity" mediation while in the second case radiative correction will become important and we call this ``loop mediation". The latter includes the much discussed scenario of ``anomaly mediation" as a special case as we shall outline in detail in section 2. There we shall provide formulae and discuss the prediction of various schemes for the soft SUSY breaking 
terms. In section 3 we shall then analyze the competing schemes and identify the signatures. Section 4 will discuss possibilities to disentangle the various schemes. In section 5 we shall set up benchmark scenarii where we analyze the 
soft terms and their robustness. Section 6 will give a conclusion and outlook.

\section{Gaugino masses in supergravity}

To start with, let us consider 4D effective supergravity (SUGRA) defined at the cutoff scale $\Lambda$ for the visible sector physics. This 4D SUGRA might correspond to the low energy limit of compactified string theory or brane model. We follow the notation of \cite{Choi:2007ka}. The Wilsonian effective action of the model at $\Lambda$ can be written as
\begin{align}
\int d^4\theta \,
C\ol{C}\Big(-3\,\e^{-K/3}\Big) + \left[ \int d^2\theta
\,\Big(\,\frac{1}{4}f_a W^{a\alpha}W^a_\alpha+ C^3 W\,\Big)+{\rm
h.c}\,\right],
\end{align}
where $C=C_0+\theta^2F^C$ is the chiral compensator of 4D SUGRA, $K$ is the K\"ahler potential, $W$ is the superpotential, and $f_a$ are holomorphic gauge kinetic functions. As usual, $K$ can be expanded as
\begin{align}
K &= K_0(X_I,\ol{X}_I)+Z_i(X_I,\ol{X}_I)\,\ol{Q}_iQ_i\,,
\end{align}
where $K_0$ denotes the moduli K\"ahler potential, $Q_i$ are chiral matter superfields which have a mass lighter than $\Lambda$ and are charged under the visible sector gauge group and $Z_i$ is the K\"ahler metric for the observable fields. $X_I$ are SUSY breaking (moduli or matter) fields which have nonzero $F$-components $F^I$.

The running gauge couplings and gaugino masses at a scale $\mu$ below $\Lambda$ but above the next threshold scale $M_{\rm th}$ can be determined by the 1PI gauge coupling superfield ${\cal F}_a(p^2)$ \mbox{($M^2_{\rm th}<p^2<\Lambda^2$)} which corresponds to the gauge kinetic coefficient in the 1PI effective action on superspace:
\begin{align}
\Gamma_{\rm 1PI} &= \int d^4p\, d^4\theta \left(\, \frac{1}{4}{\cal
F}_a(p^2) W^a\frac{{\cal D}^2}{16 p^2}W^a+{\rm h.c}\,\right).
\end{align}
In the one-loop approximation, ${\cal F}_a$ is given by \cite{Kaplunovsky:1994fg,Bagger:1999rd,ArkaniHamed:1998kj}
\begin{align}
{\cal F}_a(p^2) &= {\rm
Re}(f^{(0)}_a)-\frac{1}{16\pi^2}\klammer{3C_a-\sum_iC_a^i}\log\left(\frac{C\ol{C}\,\Lambda^2}{p^2}\right)
\nonumber \\
&~ -\frac{1}{8\pi^2}\sum_i
C_a^i\log\left(\e^{-K_0/3}Z_i\right)+\frac{1}{8\pi^2}{\Omega}_a\,,
\end{align}
where $f_a^{(0)}$ are the tree-level gauge kinetic function, $C_a$ and $C_a^i$ are the quadratic Casimir of the gauge multiplets and the matter representation $Q_i$, respectively. Here $\Omega_a$ contains the string and/or KK threshold corrections from heavy fields at scales above $\Lambda$ as well as the (regularization scheme-dependent) field-theoretic one-loop part: $\frac{1}{8\pi^2}C_a\log\big(\real (f_a^{(0)})\big)$. In the one-loop approximation, $\Omega_a$ are independent of the external momentum $p^2$, thus independent of $C$ as a consequence of the super-Weyl invariance. However $\Omega_a$ generically depend on SUSY breaking fields $X_I$, and a full determination of their $X_I$-dependence requires a detailed knowledge of the UV physics above $\Lambda$.

The running gauge couplings and gaugino masses at a renormalization point $\mu$ ($M_{\rm th}<\mu<\Lambda$) are given by
\begin{align}
\frac{1}{g_a^2(\mu)} &= {\cal F}_a|_{C=e^{K_0/6},\,p^2=\mu^2}\,,
\nonumber \\M_a(\mu) &= F^A\partial_A\log \left({\cal
F}_a\right)|_{{C=\e^{K_0/6},\,p^2=\mu^2}}\,,
\end{align}
where $F^A=(F^C,F^I)$, $\partial_A=(\partial_C,\partial_I)$, $C=\e^{K_0/6}$ corresponds to the Einstein frame condition, and
\begin{align}
\frac{F^C}{C_0} &= m^*_{3/2}+\frac{1}{3}F^I\partial_I K_0\,.
\label{compensator}
\end{align}
One then finds \cite{Kaplunovsky:1994fg,Bagger:1999rd}
\begin{align}
\nonumber \frac{1}{g_a^2(\mu)} &= {\rm Re}(f^{(0)}_a) -\frac{1}{16\pi^2}\left[\klammer{3C_a-\sum_i C_a^i}\log\left(\frac{\Lambda^2}{\mu^2}\right)\right.\\
&~ +\left.\klammer{C_a-\sum_iC_a^i}K_0+2\sum_i C_a^i\log Z_i\,\right]+\frac{1}{8\pi^2}\Omega_a\,, \\
\frac{M_a(\mu)}{g_a^2(\mu)} &= F^I\partial_I{\cal F}_a+F^C\partial_C{\cal F}_a \nonumber \\
&= F^I\left[\frac{1}{2}\partial_I f_a^{(0)}-\frac{1}{8\pi^2}\sum_i C_a^iF^I\partial_I\log\klammer{\e^{-K_0/3}Z_i}+\frac{1}{8\pi^2}\partial_I\Omega_a\right]\nonumber \\
&~ -\,\frac{1}{16\pi^2}\klammer{3C_a-\sum_iC_a^i}\frac{F^C}{C_0}\,.
\label{highscaleratio}
\end{align}
Note that $M_a/g_a^2$ do {\it not} run at one loop level, i.\,e.\ are independent of $\mu$, as $M_a$ and $g_a^2$ have the same running behavior in the one-loop approximation.

However, depending upon the SUSY breaking scenario, the ratios $M_a/g_a^2$ can receive important threshold corrections at lower intermediate threshold scales $M_{\rm th}$. In fact, the expression for $M_a/g_a^2$ in Eq.\,(\ref{highscaleratio}) is valid only for the renormalization point between the high scale $\Lambda$ and and the
intermediate scale $M_{\rm th}$ where some of the particles decouple. Let us now consider how $M_a/g_a^2$ are modified by such threshold effects at lower scale. To see this, we assume
\begin{align}
\{Q_i\}\equiv \{\Phi+\Phi^c,Q_x\}\,,
\end{align}
and $\Phi+\Phi^c$ get a supersymmetric mass of the order of $M_{\rm th}$, while $Q_x$ remain to be massless at $M_{\rm th}$. Then $\Phi+\Phi^c$ can be integrated out to derive the low energy parameters at scales below $M_{\rm th}$. The relevant couplings of $\Phi+\Phi^c$ at $M_{\rm th}$ can be written as
\begin{align}
\int d^4\theta\, C\ol{C}\, \e^{-K_0/3}\left(Z_\Phi\Phi^*\Phi+Z_{\Phi^c}\Phi^{c*}\Phi^c\right) +\left(\int d^2\theta \,C^3\lambda_\Phi X_\Phi\Phi^c\Phi+{\rm h.c}\right),
\end{align}
where $X_\Phi$ is assumed to have a vacuum expectation value\footnote{If $X_\Phi$ is not a superfield, but a parameter, then $F^{X_\Phi}$ is obviously zero.}
\begin{align}
\langle X_\Phi\rangle &= {M}_\Phi +\theta^2 F^{X_\Phi}.
\end{align}
Then the physical mass of $\Phi+\Phi^c$ are given by
\begin{align}
{\cal M}_\Phi=\lambda_\Phi\frac{C X_\Phi}{\sqrt{\e^{-2K_0/3}Z_\Phi Z_{\Phi^c}}},
\end{align}
yielding a threshold correction to the gauge coupling superfield ${\cal F}_a$ as
\begin{align}
\Delta {\cal F}_a(M_{\rm th}) &= - \frac{1}{8\pi^2}\sum_{\Phi} C_a^\Phi \log\left(\frac{{\cal M}_\Phi
 {\cal M}_\Phi^*}{M_{\rm th}^2}\right).
\end{align}
For $M_{\rm th}\sim {\cal M}_\Phi$, this gives rise to a threshold correction of ${\cal O}(1/8\pi^2)$ to $1/g_a^2$. In the leading log approximation for gauge couplings, such threshold corrections can be ignored, therefore $1/g_a^2$ obeys the continuity condition at $M_{\rm th}$:
\begin{align}
\frac{1}{g_a^2(M_{\rm th}^+)} &= \frac{1}{g_a^2(M_{\rm th}^-)}\,,
\end{align}
where $M_{\rm th}^{\pm}$ denote the scale just above/below $M_{\rm th}$. On the other hand, because $F^I$, $F^C$ and $F^{X^\Phi}$ can be quite different from each other, the threshold correction to gaugino masses at $M_{\rm th}$ can provide an important contribution to low energy gaugino masses. For $\Delta {\cal F}_a$ given above, one easily finds that the threshold correction to gaugino masses at $M_{\rm th}$ is given by
\begin{align}
M_a(M_{\rm th}^+)-M_a(M_{\rm th}^-) &= g_a^2(M_{\rm th})F^A\partial_A \Delta {\cal F}_a \nonumber\\
&= -\frac{g_a^2(M_{\rm th})}{8\pi^2} \sum_\Phi C_a^\Phi \left(\frac{F^C}{C_0}+\frac{F^{X_\Phi}}{M_\Phi}-
F^I\partial_I\log\klammer{\e^{-2K_0/3}Z_\Phi Z_{\Phi^c}}\right).
\end{align}
Adding this threshold correction to the result of (\ref{highscaleratio}), we find
\begin{align}
\left(\frac{M_a}{g_a^2}\right)_{M_{\rm th}^-} &= F^I\left[\frac{1}{2}\partial_I f_a^{(0)} -\frac{1}{8\pi^2}\sum_x
C_a^xF^I\partial_I\left(\e^{-K_0/3}Z_x\right)+\frac{1}{8\pi^2}\partial_I\Omega_a\right]\nonumber \\
&~~ +\frac{1}{8\pi^2}\sum_\Phi C_a^\Phi \frac{F^{X_\Phi}}{M_\Phi}-\frac{1}{16\pi^2}\klammer{3C_a-\sum_xC_a^x}\frac{F^C}{C_0},
\end{align}
where $\sum_x$ denotes the summation over $\{ Q_x\}$ which remain as light matter fields at $M_{\rm th}^-$.

One can repeat the above procedure, i.\,e.\ run down to the lower threshold scale, integrate out the massive fields there, and then include the threshold correction to gaugino masses until one arrives at TeV scale. Then one finally finds
\begin{align}
\left(\frac{M_a}{g_a^2}\right)_{\rm TeV}
=\widetilde{M}_a^{(0)}+ \widetilde{M}_a^{(1)}|_{\rm
loop}+\widetilde{M}_a^{(1)}|_{\rm threshold}\,,
\label{lowscaleratio}
\end{align}
where
\begin{align}
\widetilde{M}_a^{(0)} &= \frac{1}{2}F^I\partial_If_a^{(0)}\,, \label{gauginotree} \\
\widetilde{M}_a^{(1)}|_{\rm loop}
&= \frac{1}{16\pi^2}b_a\frac{F^C}{C_0}-\frac{1}{8\pi^2} \sum_mC_a^mF^I\partial_I\log\klammer{\e^{-K_0/3}Z_m}, \label{gauginoloop}\\
&= \frac{1}{16\pi^2}b_a\left( m_{3/2} + \frac{1}{3}\partial_I K_0 F^I \right)-\frac{1}{8\pi^2}
\sum_mC_a^mF^I\partial_I\log\klammer{\e^{-K_0/3}Z_m}, \label{gauginoloop2} \\
\widetilde{M}_a^{(1)}|_{\rm threshold}&= \frac{1}{8\pi^2}\sum_\Phi
C_a^\Phi\frac{F^{X_\Phi}}{M_\Phi} + \frac{1}{8\pi^2}F^I\partial_I
\Omega_a\,.
\label{gauginothreshold}
\end{align}
Here $\sum_m$ denotes the summation over the light matter multiplets $\{Q_m\}$ at the TeV scale, $\sum_\Phi$ denotes the summation over the gauge messenger fields $\Phi+\Phi^c$ which have a mass lighter than $\Lambda$ but heavier than TeV, and
\begin{align}
b_a &= -3C_a+\sum_m C_a^m\,, \quad b_a=\left\lbrace \frac{33}{5},\,1,\,-3 \right\rbrace,
\label{betafunc}
\end{align}
are the one-loop $\beta$-function coefficients at TeV. Obviously, $\widetilde{M}_a^{(0)}$ is the tree level value of $M_a/g_a^2$, $\widetilde{M}_a^{(1)}|_{\rm loop}$ denotes contributions from radiative corrections $\widetilde{M}_a^{(1)}|_{\rm threshold}$ are threshold corrections due to intermediate and/or heavy states.

Depending upon the SUSY breaking scenario, $M_a/g_a^2$ are dominated by some of these five contributions. The threshold corrections encoded in $\widetilde{M}_a^{(1)}|_{\rm threshold}$ are most difficult to compute and highly model-dependent. In fact, they represent potentially uncontrollable contributions from high energy modes. If this part
gives an important contribution to $M_a/g_a^2$ model independent statements about the gaugino masses will be impossible. The other parts, however, can be computed reliably within the framework of the 4D effective supergravity theory in many SUSY breaking scenarii. 

Eqs. (\ref{lowscaleratio})--(\ref{gauginothreshold}) give the most general description of gaugino masses and connect them to the properties of the underlying schemes. This is our basic tool to analyze potential candidates from future collider experiments. The SM gauge coupling constants at the Tera scale can be deduced from the experimental values at $M_Z$. This gives (approximately):
\begin{align}
g_1^2:g_2^2:g_3^2\simeq 1:2:6\,.
\label{couplingsratio}
\end{align}
Once the gaugino mass ratios at TeV are measured, the ratios of $M_a/g_a^2$ at TeV can be experimentally determined and crosschecks for unification can be analyzed.

\section{Controllable mass patterns}

In the search for a controllable mass pattern we are obliged to examine simple schemes, where unification is not obscured by intermediate thresholds. Thresholds as given in Eq.\,(\ref{gauginothreshold}) at an intermediate and/or
high scale are highly model dependent. They could arise in the framework of models of gauge mediation or string thresholds at the large scale and spoil the predictions. Of course, in some models these thresholds might conspire and cancel each other. In this case the prediction of the simple model might not be altered and we just consider the simple model by itself. The notion of simple schemes includes the notion of simple boundary conditions at the GUT scale. We shall here concentrate
on universal parameters.

\subsection{Gravity Mediation}

The scheme of gravity mediation \cite{Nilles:1982ik,Chamseddine:1982jx,Ibanez:1982ee,Barbieri:1982eh,Nilles:1982dy,Hall:1983iz} with a universal gaugino mass at $M_{\rm GUT}$ is the one of the most popular scenarii whose phenomenological consequences have been studied extensively under the name of mSUGRA scenario. In this scheme, $M_a$ are determined by the gravitino mass and are thus assumed to be universal at $M_{\rm GUT}$, leading to the minimal supergravity (mSUGRA) pattern of gaugino masses at the TeV scale\footnote{For the numerical evaluation we use a renormalization group scale of $\mu = \unit[500]{GeV}$.}:
\begin{align}
M_1:M_2:M_3\,\simeq\, 1: 2: 6\,.
\label{msugrapattern}
\end{align}
In the language of 4D effective SUGRA discussed in the previous section, this amounts to assuming that the cutoff scale $\Lambda$ of 4D effective SUGRA is close to $M_{\rm P}$ or $M_{\rm GUT}$, and $M_a/g_a^2$ of Eq.\,(\ref{lowscaleratio}) are dominated by the contribution determined by the tree-level gauge kinetic function: $\widetilde{M}_a^{(0)}=\frac{1}{2}F^I\partial_If_a^{(0)}$, which is assumed (or predicted) to be universal. Some interesting examples of such scenario include dilaton mediation in heterotic string/$M$-theory \cite{Nilles:1998sx,Lukas:1997fg,Choi:1997cm}, flux-induced SUSY breakdown in Type IIB string theory \cite{Grana:2003ek,Lust:2004fi,Camara:2004jj,Lust:2004dn,Font:2004cx}, and gaugino mediation realized in brane models \cite{Kaplan:1999ac,Chacko:1999mi,Schmaltz:2000gy}. 

The soft scalar square masses and the $A$-terms at the GUT scale are \cite{Brignole:1993dj,Brignole:1997dp}
\begin{align}
A_{ijk} &= F^I\partial_I K_0 + F^I\partial_I\log\left(\frac{Y_{ijk}}{Z_i Z_j Z_k}\right),  \\
m^2_i &= \frac{2}{3}V_0 + \frac{1}{3}\overline{F}^{\overline m} F^n \partial_{\overline m}\partial_n K_0 - \frac{1}{3}\overline{F}^{\overline m} F^n \partial_{\overline m}\partial_n \log Z_i\,,
\end{align}
where $m$, $n$ run over SUSY breaking fields, $Y_{ijk}$ denote the holomorphic Yukawa couplings and $Z_i$ is the K\"ahler metric of the visible fields. In the mSUGRA scheme one assumes a canonical K\"ahler potential and minimal kinetic terms in the supergravity Lagrangian. Taking $Y_{ijk}$ to be independent of the supersymmetry breaking fields one obtains universal soft terms. Universal boundary conditions at the GUT scale provide a reason for the absence of a flavor problem. In explicit schemes the universality of the soft terms should come as a consequence of some (discrete) symmetries \cite{Kobayashi:2006wq,Ko:2007dz,Lebedev:2007hv,Nilles:2008gq}.

It is clear that the pattern for the gaugino masses as given in Eq.\,(\ref{msugrapattern}) is a clean and convincing signal for unification. The results for the soft scalar masses are more model dependent. The pattern of the soft scalar masses thus might be used as a backup to distinguish between various models of that type. Simple schemes will give 
definite relations between squark, slepton and gaugino masses as will be discussed in the next chapter.

\subsection{Loop Mediation}

So far we have considered schemes where the gaugino masses are nonzero at tree level. But there are several models, where the gaugino masses vanish at that level. Then radiative corrections Eq.\,(\ref{gauginoloop}) become relevant. We call this scheme ``Loop Mediation''. It is a scheme where the coupling between the observable and hidden sector is even weaker than in gravity mediation. There are two contributions in Eq.\,(\ref{gauginoloop}), and we discuss them separately.

\subsubsection{Anomaly Mediation}

In anomaly mediation \cite{Randall:1998uk,Giudice:1998xp}, SUSY breaking fields $X_I$ are assumed to be (effectively) sequestered from the visible sector fields where the only contribution to the gaugino masses is due to contribution mediated by the SUGRA compensator \cite{Randall:1998uk,Luty:2000ec,Kobayashi:2001kz,Luty:2001jh,Ibe:2005pj,Schmaltz:2006qs}. Due to the sequestering the SUGRA compensator Eq.\,(\ref{compensator}) is described by $F^C/C_0=m_{3/2}$. Thus the anomaly mediated contribution is the first term in Eq.\,(\ref{gauginoloop2}). Then $M_a/g_a^2$ are determined as
\begin{align}
\left(\frac{M_a}{g_a^2}\right)_{\rm
TeV}\,\simeq\,\widetilde{M}_a^{(1)}|_{\rm conformal}=\frac{b_a}{16\pi^2}\, m_{3/2}\,.
\end{align}
The relevant quantities in this case are the coefficients of the $\beta$-functions. If the effective theory around TeV is given by the MSSM, Eq.\,(\ref{betafunc}) and the low energy gaugino masses take the anomaly pattern: 
\begin{align}
M_1:M_2:M_3 \,\simeq\, 3.3:1:9\,.
\label{anomalypattern}
\end{align}
Again this is a very simple and predictive scheme. An observation of the above pattern would be a spectacular result.

The soft scalar masses squared and the $A$-terms in anomaly mediation are determined by the MSSM $\gamma$-functions
\begin{align}
A_{ijk} &= \frac{\gamma_i+\gamma_j+\gamma_k}{16\pi^2}\, m_{3/2}, \\
m^2_i &= -\frac{\dot{\gamma}_i}{(16\pi^2)^2}\left|m_{3/2}\right|^2,
\end{align}
where $\gamma_i$ denotes the anomalous dimension of $Q_i$ and $\dot{\gamma}_i=8\pi^2\frac{\del\gamma_i}{\del\log\mu}$. For details see \cite{Choi:2004sx,Choi:2005ge}. In the MSSM $\dot{\gamma}_i$ is positive for the sleptons, indicating that they are tachyonic at the GUT scale. Thus the scheme of anomaly mediation requires additional contributions to the scalar masses. Then the question remains whether this mechanism affects the gaugino mass relation Eq.\,(\ref{anomalypattern}) as well. We shall come back to this question later. Up to now, we have only considered that part in Eq.\,(\ref{gauginoloop2}) that appears in the case of complete sequestering. In general we also have to take into account the other loop contributions.

\subsubsection{K\"ahler Mediation}

We now consider the remaining contributions
\begin{align}
\widetilde{M}_a^{(1)}|_{\rm K\ddot{a}hler} &= \frac{b_a\, g^2_a}{16\pi^2}F^I \partial_I\frac{K_0}{3}-\frac{1}{8\pi^2}
\sum_m C_a^m F^I\partial_I\log\klammer{\e^{-K_0/3}Z_m},
\label{gauginokahler}
\end{align}
which depends on the K\"ahler potential connecting the SUSY breakdown sector to the observable sector. This is a contribution beyond that of complete sequestering (as given in anomaly mediation) and it potentially destroys the beautiful mass pattern of the anomaly mediation scheme. We parameterize this ambiguity by a parameter $\phi$ that represents vacuum expectation values of the unsuppressed (partially sequestered) SUSY breaking fields in the hidden sector. Given a minimal form of the K\"ahler metric for the visible fields $Z_i$, Eq.\,(\ref{gauginokahler}), reduces to
\begin{align}
\widetilde{M}_a^{(1)}|_{\rm K\ddot{a}hler}
=\frac{b^\prime_a\, g^2_a}{16\pi^2}F^I \partial_I K_0\,,
\end{align}
with 
\begin{align}
b^\prime_a &= -C_a + \sum_m C^m_a\,, \quad b^\prime_a=\left\lbrace \frac{33}{5},\,5,\,3 \right\rbrace.
\label{betaprime}
\end{align}
Fortunately these coefficients are completely determined by the low energy spectrum of the MSSM. Thus the contribution to loop mediation consists of two terms, one term proportional to $b_a$, Eq.\,(\ref{betafunc}), and the other proportional to $b^\prime_a$. With K\"ahler Mediation we denote the contribution proportional to $b_a^\prime$. In its pure form it would lead to a low energy gaugino mass pattern 
\begin{align}
M_1:M_2:M_3 &\simeq 3.3:5:9\,,
\label{kahlerpattern}
\end{align}
again a distinctive relation. This is a contribution beyond that of complete sequestering (represented by anomaly mediation). K\"ahler mediation might suffer from the tachyon problem as well. But again this will be a question concerning soft breaking parameters that are more model dependent than the gaugino masses.

On the other hand the contribution of K\"ahler mediation to the gaugino masses might render our analysis difficult and potentially inconclusive. We have to see how we can disentangle the various contributions. 

With loop mediation we denote the sum of anomaly and K\"ahler mediation. The gaugino mass ratios at the TeV scale are
\begin{align}
M_1 : M_2 : M_3 &\simeq |3.3+3.3\,\phi| : |1+5\,\phi| : |-9+9\,\phi|\,.
\label{looppattern}
\end{align}
In general, loop mediation might have problems with potential tachyons and and one would need new contributions for the soft scalars to avoid this problem. While this problem could be solved in various ways with or without disturbing the gaugino mass pattern, a natural way to avoid tachyons would be a nonzero tree level contribution
to the soft mass terms. This would lead us to a scheme where gravity mediation and loop mediation are both relevant.

\subsection{Mirage mediation}

This is a scheme which in its pure form \cite{Choi:2004sx,Choi:2005ge} is a superposition of gravity and anomaly mediation. For the moment we shall assume that the contribution from K\"ahler mediation is absent or subdominant. The name mirage comes from the fact that in such a scheme the gaugino masses unify at the so-called mirage scale, somewhere in between the Tera scale and the GUT scale. In that sense the footprints of unification will show up via the reconstruction of the
mirage scale from LHC data.

The characteristic feature of the mirage mediation scheme is the appearance of the so-called \emph{little hierarchy} $\log(M_{\rm P}/m_{3/2})$ \cite{Choi:2004sx,Choi:2005ge,LoaizaBrito:2005fa}. The little hierarchy acts to suppress the contribution from gravity mediation 
\begin{align}
F^{\mathrm{moduli}} &\sim \frac{m_{3/2}}{\log{(M_{\rm P}/m_{3/2})}}\,.
\end{align}
For a TeV gravitino mass the logarithm is comparable to a 1-loop suppression factor. Thus SUSY breaking contributions from loop mediation become competitive. In its minimal form mirage mediation contains only the contribution from moduli/gravity and anomaly mediations of comparable size, i.\,e. $\phi=0$. In the scheme of matter up-/down-lifting \cite{GomezReino:2006dk,Lebedev:2006qq,Dudas:2006gr,Abe:2006xp,Lebedev:2006qc,Lowen:2008fm} complete sequestering is generically absent and the auxiliary components of matter fields are no longer suppressed. Therefore the contribution from K\"ahler mediation Eq.\,(\ref{gauginokahler}) becomes relevant. The gaugino 
masses at the GUT scale can be expressed as 
\begin{eqnarray}
M_a &= m_{1/2} \Bigg[ 1 + \alpha\,\frac{\log(M_{\rm P}/m_{3/2})}{16\pi^2}\,b_a\,g^2_a + \alpha\,\frac{\log(M_{\rm P}/m_{3/2})}{16\pi^2}\,\phi\,b^\prime_a\,g^2_a \Bigg],
\end{eqnarray}
where we have introduced
\begin{align}
\alpha &= \frac{m_{3/2}}{m_{1/2}\log(M_{\rm P}/m_{3/2})}\,, \\
m_{1/2} &= F^I\partial_I\log\klammer{\real\big(f^{(0)}_a\big)}.
\end{align}
The parameter $\alpha$ measures the ratio between modulus and anomaly mediation. $\alpha=0$ corresponds to pure 
modulus/gravity mediation whereas for $\alpha\rightarrow\infty$ (with $\phi=0$) one obtains pure anomaly mediation.

Suppose $\phi=0$. We denote this case as \emph{pure} mirage mediation. The gaugino mass ratios at the GUT scale 
are then given by the respective $\beta$-function coefficients $b_a$. Since the RG evolution is governed by the same $\beta$-functions, the splitting of the gaugino masses cancels at an intermediate scale. At 1-loop level one obtains the mirage scale \cite{Choi:2005uz}
\begin{align}
M_{\mathrm{MIR}} &= M_{\mathrm{GUT}}\left(\frac{m_{3/2}}{M_{\mathrm{P}}}\right)^{\alpha/2}.
\label{miragescale}
\end{align}
\begin{figure}
\captionsetup[figure]{labelfont={footnotesize,bf},textfont=footnotesize,labelsep=mysep,labelformat=mypiccap,format=default,justification=RaggedRight,width=7cm,indent=5pt}
\begin{minipage}{0.5\linewidth}
\includegraphics[width=\linewidth]{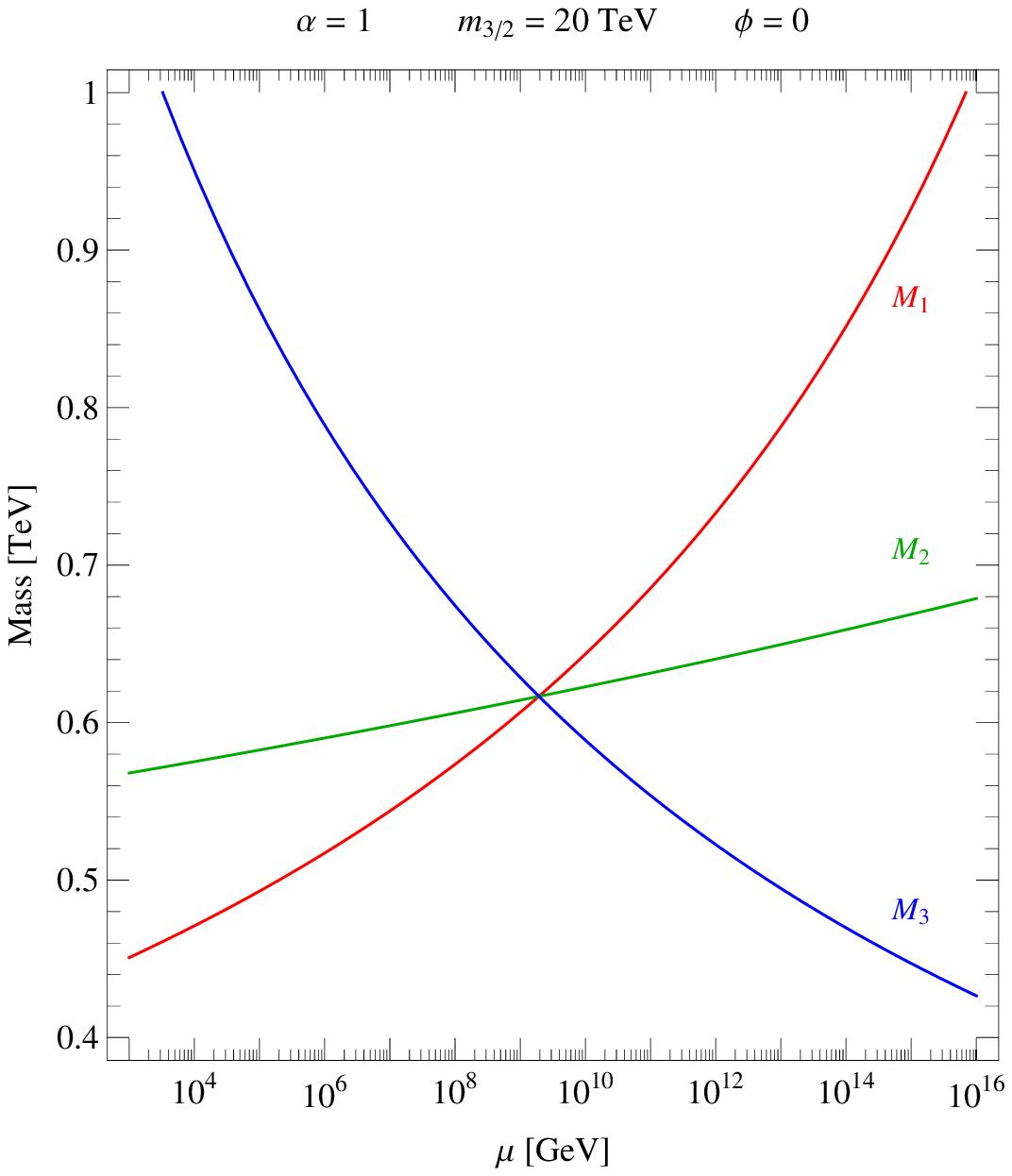}
\caption{Mirage unification of the gaugino masses in the scheme of pure mirage mediation.}
\label{fig:mirage}
\end{minipage}
\begin{minipage}{0.5\linewidth}
\includegraphics[width=\linewidth]{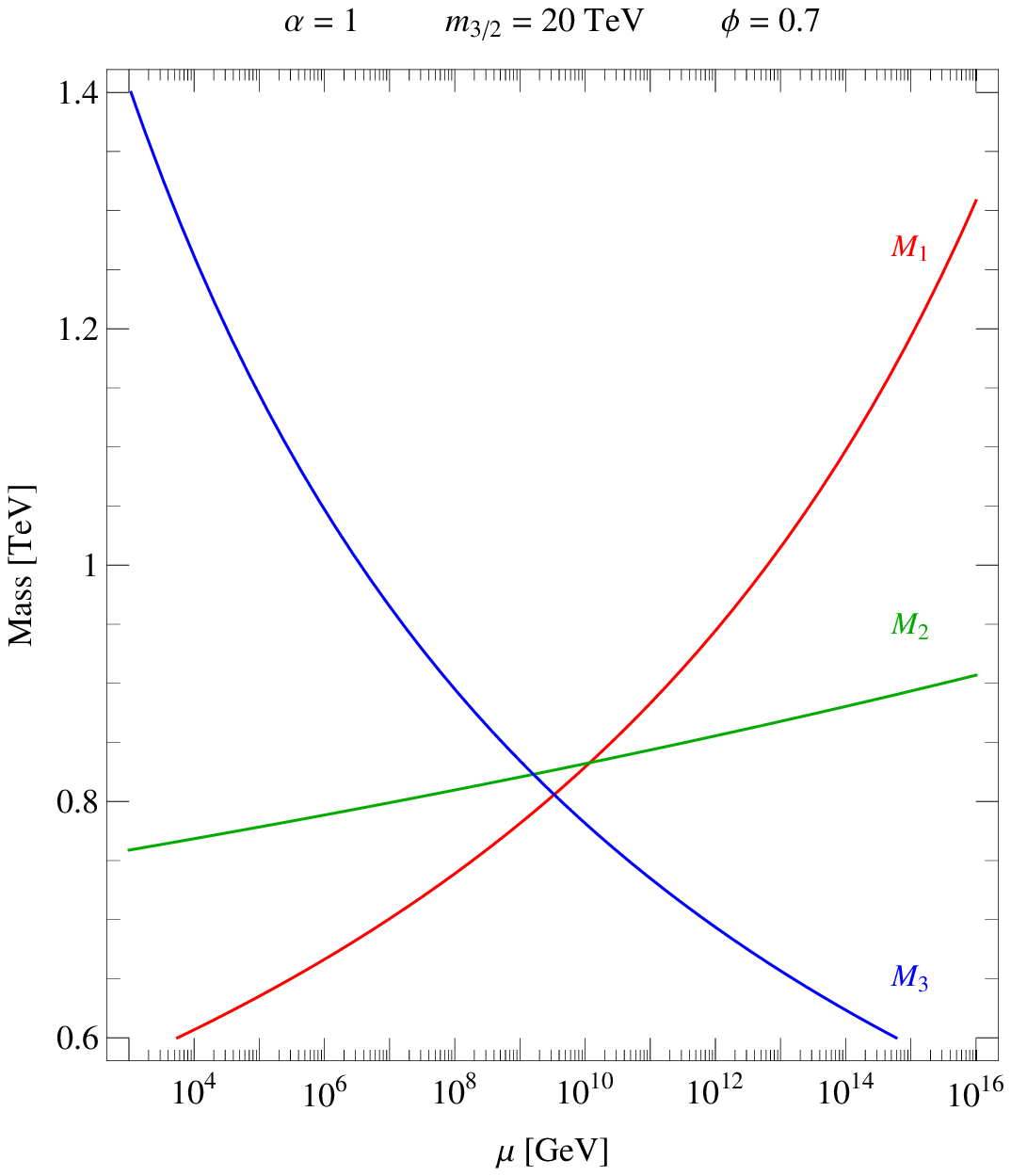}
\caption{Deviation from mirage unification caused by the non-universal contribution from K\"ahler mediation.}
\label{fig:kahler}
\end{minipage}
\end{figure}
\begin{figure}
\captionsetup[figure]{labelfont={footnotesize,bf},textfont=footnotesize,labelsep=mysep,labelformat=mypiccap,format=default,justification=RaggedRight,width=7cm,indent=5pt}
\begin{minipage}{0.5\linewidth}
\includegraphics[width=\linewidth]{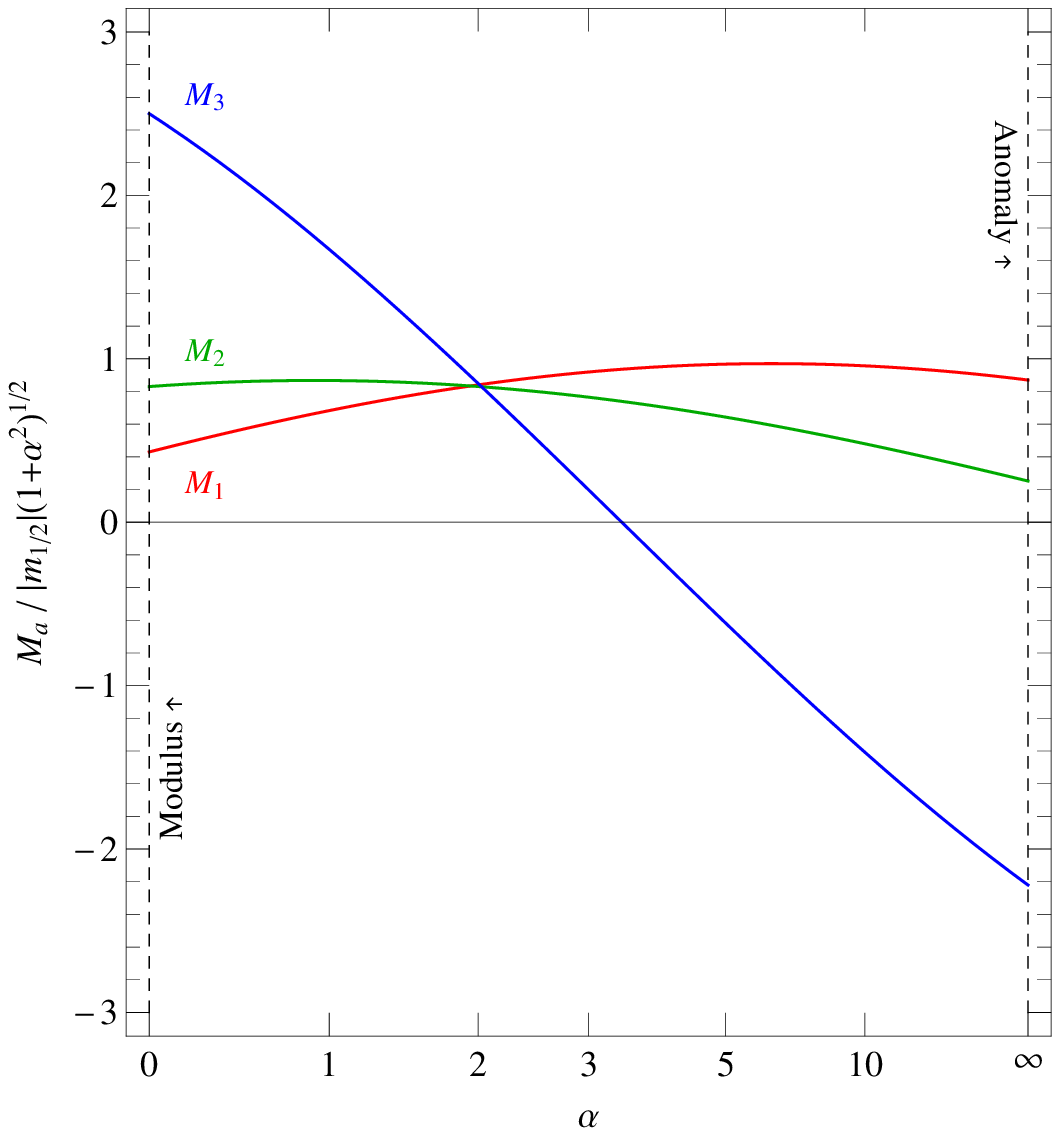}
\caption{Gaugino masses at the Tera scale as a function of $\alpha$.}
\label{fig:gauginos}
\end{minipage}
\begin{minipage}{0.5\linewidth}
\includegraphics[width=\linewidth]{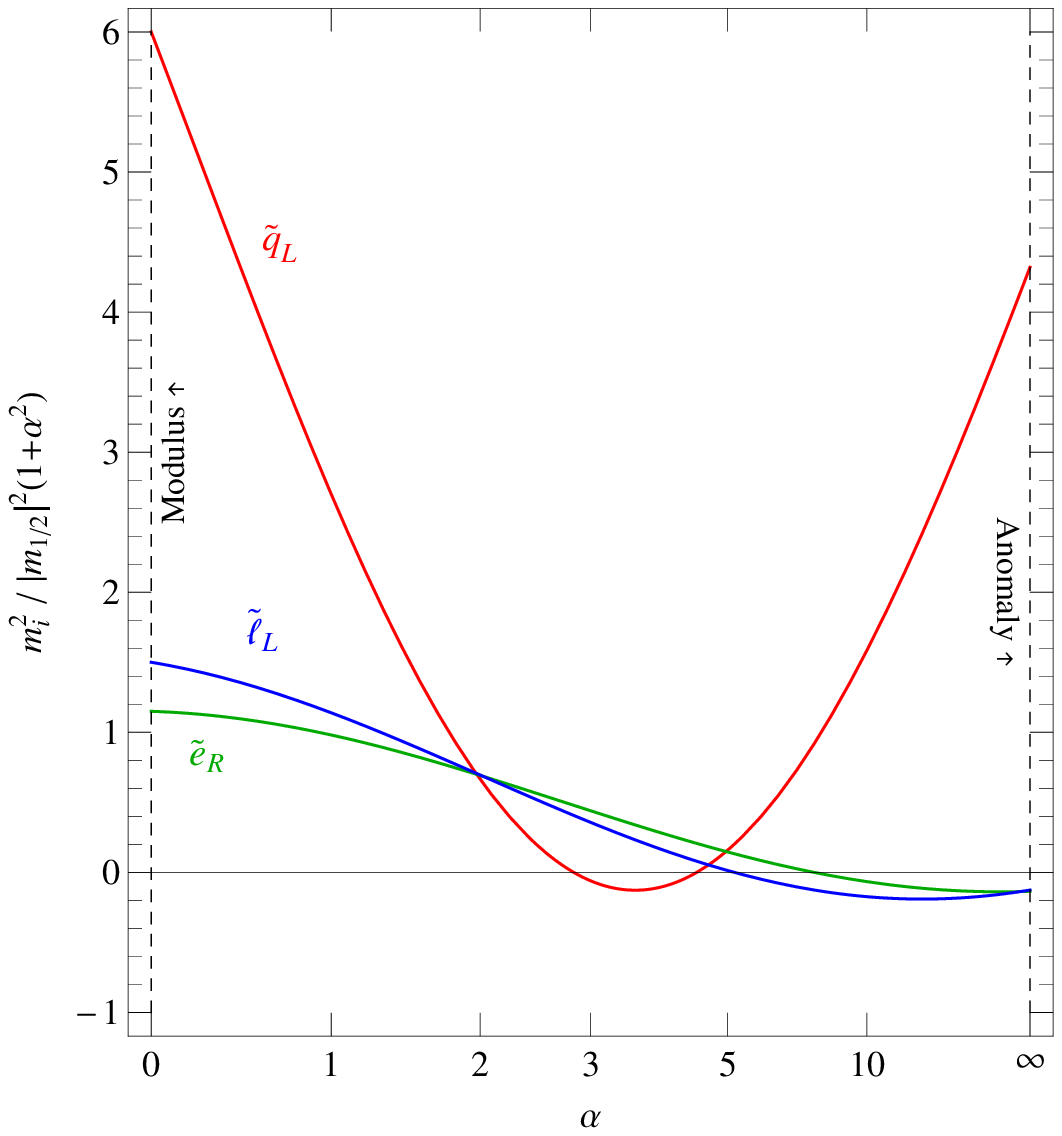}
\caption{Scalar squared masses at the Tera scale as a function of $\alpha$ with $m_0=m_{1/2}$.}
\label{fig:scalars}
\end{minipage}
\end{figure}
This scale does not correspond to a physical threshold, hence the name mirage. For $\alpha=1$ mirage unification occurs at an intermediate scale between the GUT and the TeV scale. For this value of $\alpha$ the evolution of the gaugino masses is illustrated in Fig.\,\ref{fig:mirage}. An interesting feature of mirage mediation is that due to the large negative $\beta$-function of the SU(3), the contribution from anomaly mediation considerably lowers the gluino mass 
at the GUT scale. Moreover for $\alpha\sim2$ the mirage scale would coincide with the Tera scale. Fig.\,\ref{fig:gauginos} illustrates the $\alpha$-dependence of the gaugino masses as the Tera scale.

Allowing $\phi\neq0$ will introduce further non-universal contributions to the gaugino masses which are proportional to $b^\prime_a$ that potentially destroy the mirage unification feature (cf.\ Fig.\,\ref{fig:kahler}), although we might hope that the violation of the mirage behavior could still be rather mild.

At 1-loop level the full contribution to the gaugino masses is given by
\begin{align}
M_1 : M_2 : M_3 &\simeq |1+0.66\,\alpha+0.66\,\alpha\,\phi| : |2+0.2\,\alpha+\alpha\,\phi| : |6-1.8\,\alpha+1.8\,\alpha\,\phi|\,,
\label{generalpattern}
\end{align}
which is controlled by the two parameters $\alpha$ and $\phi$. An interesting limit is $\alpha=\phi=1$. In this case loop mediated contributions to the gluino mass cancel.

\section{Disentangling the schemes}

We have seen that even under our simplified assumptions it might become difficult to determine the parameters. This is the reason why we refrain from a discussion of more complicated schemes and boundary conditions. Things might be easy if one of the schemes is realized in a more or less pure form. Otherwise a careful analysis is needed. In order to disentangle the schemes other soft parameters are needed. In a first step we might look at the ratios of squark/slepton and gaugino masses at the TeV-scale.\footnote{For the numerical evaluation we use
the renormalization group scale $\mu = \unit[500]{GeV}$.} 
We assume universal mass parameters at the GUT scale: $m_{1/2}$ for the gauginos and $m_0$ for the scalar partners of quarks an leptons. For the first two generations one finds (at the one-loop level)) \cite{Choi:2009jn}
\begin{align}
\nonumber m^2_{\widetilde{q}_L} &= m^2_0 + \Big[ 5 -  3.48\,\alpha  + 0.48\,\alpha^2 \Big] m_{1/2}^2\,, \\
\nonumber m^2_{\widetilde{u}_R} &= m^2_0 + \Big[ 4.6 -  3.29\,\alpha  + 0.49\,\alpha^2 \Big] m_{1/2}^2\,, \\
m^2_{\widetilde{d}_R} &= m^2_0 + \Big[ 4.5 -  3.27\,\alpha  + 0.49\,\alpha^2 \Big] m_{1/2}^2\,, \label{lowmasses}\\
\nonumber m^2_{\widetilde{l}_L} &= m^2_0 + \Big[ 0.5 -  0.22\,\alpha  - 0.014\,\alpha^2 \Big] m_{1/2}^2\,, \\
\nonumber m^2_{\widetilde{e}_R} &= m^2_0 + \Big[ 0.15 - 0.045\,\alpha - 0.015\,\alpha^2 \Big] m_{1/2}^2\,. 
\end{align}
In Fig.\,\ref{fig:scalars} we depict the scalar squared masses at the Tera scale for the case $m_0=m_{1/2}$.

\subsection{Gravity mediation}

The observation of a gaugino mass pattern 
\begin{align}
M_1:M_2:M_3\,\simeq\, 1: 2: 6\,.
\end{align}
would be a spectacular result towards a confirmation of a grand unified picture. Since in this case a bino-like LSP seems to be rather generic the gluino to LSP ratio would then be given by
\begin{align}
\mathscr{G}=\frac{M_3}{\widetilde{\chi}^0_1}\simeq 6\,.
\end{align}
Its simplest manifestation is in the scheme of gravity/modulus mediation with universal boundary conditions, but there might be other schemes that give a similar answer. 

Having $\alpha=0$ Eq.\,(\ref{lowmasses}) gives us the squark and slepton masses in gravity mediation
\begin{align}
\nonumber m^2_{\widetilde{q}_L} &= m^2_0 + 5\,  m_{1/2}^2\,, \\
\nonumber m^2_{\widetilde{u}_R} &= m^2_0 + 4.6\,  m_{1/2}^2\,, \\
m^2_{\widetilde{d}_R} &= m^2_0 + 4.5\,  m_{1/2}^2\,, \\
\nonumber m^2_{\widetilde{l}_L} &= m^2_0 + 0.5\,  m_{1/2}^2\,, \\
\nonumber m^2_{\widetilde{e}_R} &= m^2_0 + 0.15\, m_{1/2}^2\,.
\end{align}
It is clear that in such a scheme squarks will be significantly heavier than sleptons, but a quantitative analysis requires great care and precision. Furthermore, there might be uncertainties which one has to take into account like e.\,g.\ left-right splitting of the masses and also the question of running versus physical masses \cite{Choi:2009jn}.

\begin{figure}
\captionsetup[figure]{labelfont={footnotesize,bf},textfont=footnotesize,labelsep=mysep,labelformat=mypiccap,format=default,justification=RaggedRight,width=7cm,indent=5pt}
\begin{minipage}{0.5\linewidth}
\includegraphics[width=\linewidth]{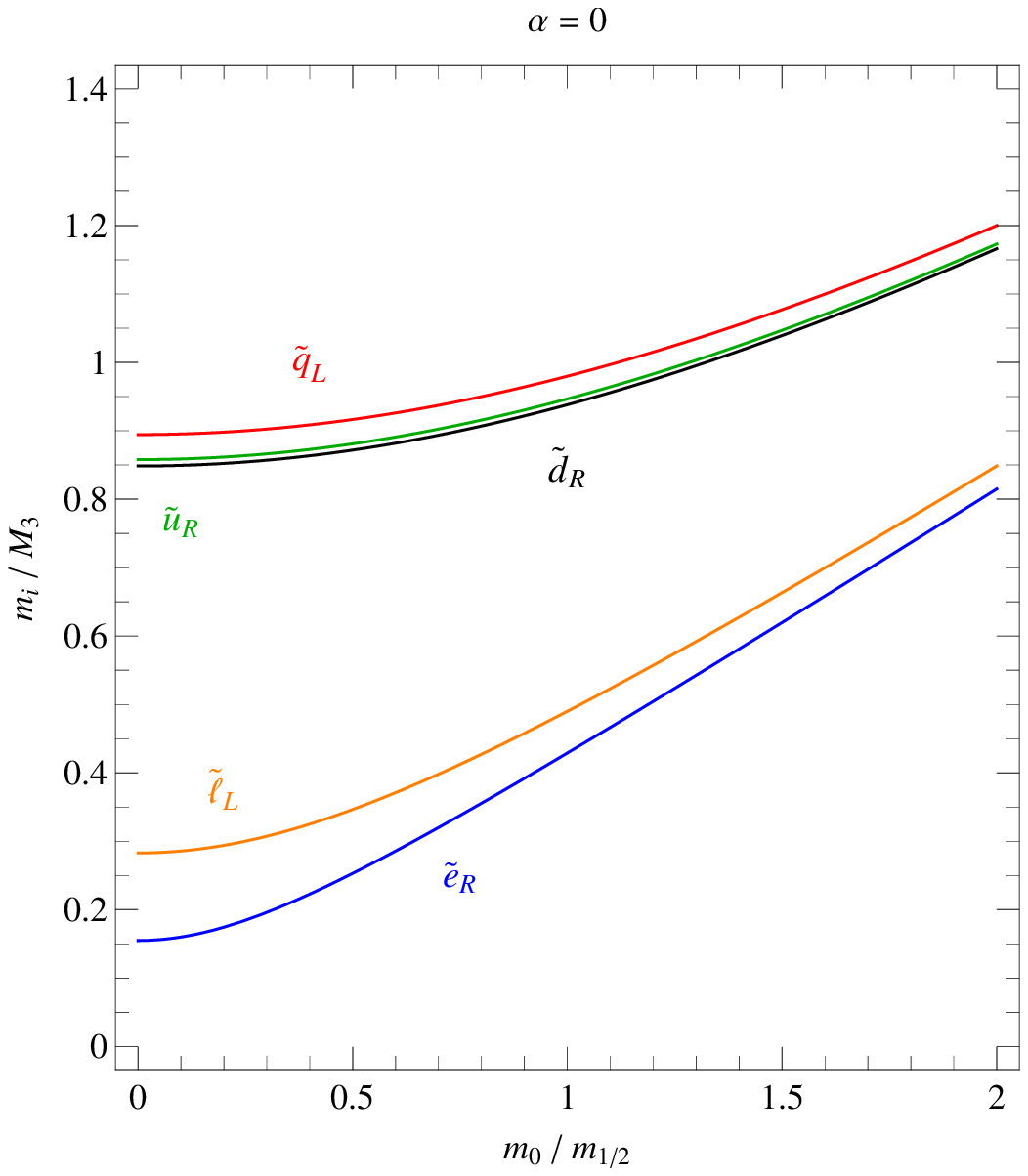}
\caption{Squark to gluino mass ratio at the Tera scale in pure gravity/modulus mediation as a function of the ratio $m_0/m_{1/2}$.}
\label{fig:msugrarelation}
\end{minipage}
\begin{minipage}{0.5\linewidth}
\includegraphics[width=\linewidth]{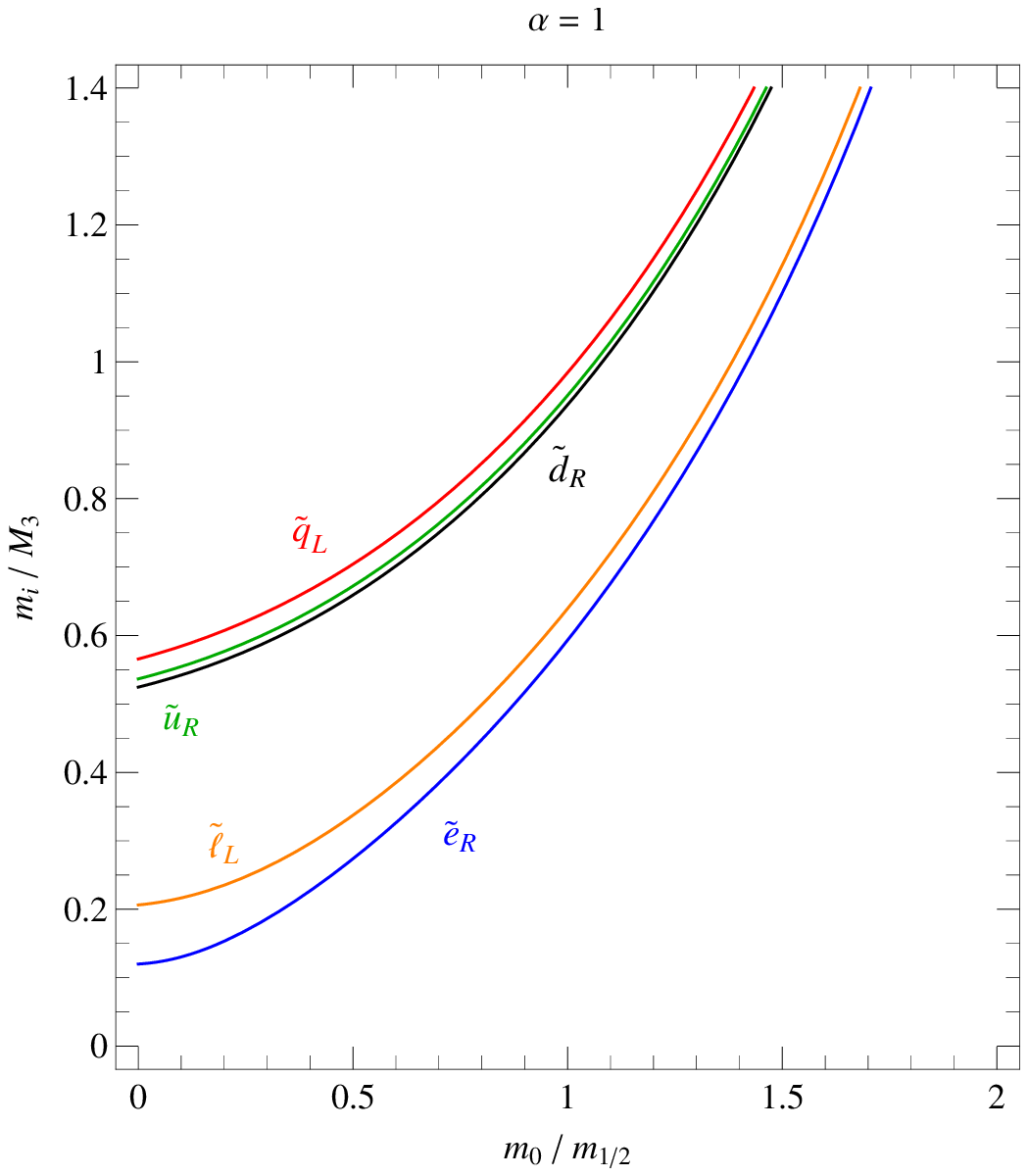}
\caption{Squark to gluino mass ratio at the Tera scale in pure mirage mediation as a function of the ratio $m_0/m_{1/2}$.}
\label{fig:miragerelation}
\end{minipage}
\end{figure}

In any case, we shall also compare squark to gluino masses to check for universality of soft masses at the GUT scale. In case of $m_0=m_{1/2}$ the squark to gluino mass ratio would be
\begin{align}
\frac{m_{\widetilde{q}_L}}{M_3}\approx 0.98\,.
\end{align}
The general case for the ratio of the scalar masses is shown in Fig.\,\ref{fig:msugrarelation}.
While the $A$-parameter is theoretically easier to determine than the soft scalar masses, experimentally its determination is a challenge. Its simplest manifestation might appear in a study of the left-right splitting of the stop squark and it is not clear whether the LHC can give us the precision to determine the $A$-parameters. Apart from this question of the $A$-parameters we are convinced that the simplest scheme of gravity/modulus mediation could be verified at the LHC.

\subsection{Loop mediation}

Another characteristic pattern of the gaugino masses is given by 
\begin{align}
M_1:M_2:M_3 \,\simeq\, 3.3:1:9\,.
\label{anomalypattern2}
\end{align}
It is a spectacular signal for the anomaly mediation scheme. This is quite special as it is the only simple scheme that favors a wino-like LSP. In addition the gluino to LSP mass ratio is practically as high as it can be:
\begin{align}
\mathscr{G}=\frac{M_3}{\widetilde{\chi}^0_1}\simeq 9\,.
\end{align}
Setting $\alpha\rightarrow\infty$ in Eq.\,(\ref{lowmasses}) gives us the soft scalar masses in pure anomaly mediation. These scalar masses show a distinct pattern
\begin{align}
\nonumber m^2_{\widetilde{q}_L} &=   0.48\, m_{1/2}^2\,, \\
\nonumber m^2_{\widetilde{u}_R} &=   0.49\, m_{1/2}^2\,, \\
m^2_{\widetilde{d}_R} &=   0.49\, m_{1/2}^2\,, \\
\nonumber m^2_{\widetilde{l}_L} &= -0.014\, m_{1/2}^2\,, \\
\nonumber m^2_{\widetilde{e}_R} &= -0.015\, m_{1/2}^2\,.
\end{align}
Unfortunately the pure anomaly mediation scheme exhibits tachyonic sleptons and one needs new contributions to the scalar masses that remove the tachyons. One might hope that this mechanism does not influence the gaugino masses nor the squark masses substantially.

In anomaly mediation we have for the ratio of squark to gluino masses
\begin{align}
\frac{m_{\widetilde{q}_L}}{M_3}\approx 0.94\,.
\end{align}
Independently of these uncertainties of the scalar masses the observation of the relation Eq.\,(\ref{anomalypattern2}) would be a conclusive signal of anomaly mediation.

Pure K\"ahler mediation would lead to the pattern Eq.\,(\ref{kahlerpattern}) and is not as distinctive as the pattern of pure gravity and anomaly mediation. The scheme prefers a bino-like LSP but the gluino to LSP mass ratio is rather small:
\begin{align}
\mathscr{G}=\frac{M_3}{\widetilde{\chi}^0_1}\simeq 2.7\,.
\end{align}
One would have to invest some effort to distinguish it from mixed schemes as e.\,g.\ mirage mediation. It is difficult to give a meaningful general formula for soft scalar squared masses as they might depend on more parameters than just the value of $\phi$. We expect again a potential problem with tachyons. For the gaugino masses one might observe that in the scheme of pure K\"ahler mediation the combination $M_1+M_2-M_3$ tends to be rather small.

In addition, it might not appear in its pure form since as anomaly mediation the K\"ahler scheme relies on radiative corrections and one might expect a mixed form: loop mediation. Due to the cancellation in the gluino mass in Eq.\,(\ref{looppattern}) we expect a rather homogeneous mass pattern. All three values turn out to be rather close to each other. For $\phi$ close to $1$ we have a very light gluino, but this is problematic in realistic models. In any case it might be difficult to pin down loop mediation for arbitrary values of the mixing parameter $\phi$.
One should mention here that the contribution of K\"ahler mediation might be forbidden by a symmetry in a large class of models. Thus one could hope that the simpler scheme of anomaly mediation might be realized in nature.

\subsection{Mirage mediation}

The gaugino mass pattern in pure mirage mediation
\begin{align}
M_1 : M_2 : M_3 &\simeq |1+0.66\,\alpha| : |2+0.2\,\alpha| : |6-1.8\,\alpha|\,,
\end{align}
depends on the parameter $\alpha$ that determines the admixture of anomaly to gravity mediation. This parameter is determined through properties of the underlying (string) theory. In some theories $\alpha$ might be quantized.

In the first example we found $\alpha=1$. In this case the mirage scale is intermediate between the GUT and the Tera scale. For $\alpha=2$ the mirage scale is as low as a TeV. This could alleviate the so-called ``MSSM hierarchy problem'' \cite{Choi:2005uz,Choi:2005hd,Lebedev:2005ge,Pierce:2006cf} but one would have to see whether this scheme can lead to a realistic phenomenology.

As an illustrative example we consider here $\alpha=1$. The gaugino mass ratios are 
\begin{align}
M_1 : M_2 : M_3 &\simeq 1:1.3:2.5\,,
\end{align}
a rather compressed mass pattern. Again this signature is not as distinct as that for the gravity/modulus or anomaly scheme and looks quite similar to the K\"ahler scheme Eq.\,(\ref{kahlerpattern}). To confirm the mirage scheme we need to investigate the scalar mass pattern. It is given by
\begin{align}
\nonumber m^2_{\widetilde{q}_L} &= m^2_0 + 2\, m_{1/2}^2\,,    \\
\nonumber m^2_{\widetilde{u}_R} &= m^2_0 + 1.8\, m_{1/2}^2\,,  \\
m^2_{\widetilde{d}_R} &= m^2_0 + 1.72\, m_{1/2}^2\,, \\
\nonumber m^2_{\widetilde{l}_L} &= m^2_0 + 0.266\, m_{1/2}^2\,, \\
\nonumber m^2_{\widetilde{e}_R} &= m^2_0 + 0.09\, m_{1/2}^2\,,
\end{align}
while the squark to gluino mass ratio (for $m_0=m_{1/2}$) would read
\begin{align}
\frac{m_{\widetilde{q}_L}}{M_3}\approx 0.98\,,
\end{align}
with the general case being shown in Fig.\,\ref{fig:miragerelation}. Although this looks very similar to the pure gravity/modulus scheme, with the help of the scalar masses we might nevertheless be able to
identify the mirage scheme unambiguously. In the pure mirage mediation picture the LSP is predominantly bino-like leading to
\begin{align}
\mathscr{G}=\frac{M_3}{\widetilde{\chi}^0_1}\simeq 2.5\,,
\end{align}
which as mentioned in the previous subsection looks very similar to the pure K\"ahler scheme. 

A further strategy would be to design special sum rules along the lines discussed in \cite{Choi:2007ka}. In the scheme of pure gravity and anomaly mediation e.\,g.\ the value of $2(M_1+M_2)-M_3$ is rather small. In this way a non-vanishing value of this combination can be used for a possible identification of the
mirage scheme.

\subsection{Other mixed mediation schemes}

In a next step we might consider the mixture of gravity/modulus and K\"ahler mediation. The low energy gaugino mass pattern becomes
\begin{align}
M_1 : M_2 : M_3 &\simeq |1+0.66\,\phi| : |2+\,\phi| : |6+1.8\,\phi|\,.
\end{align}
Compared to pure gravity/modulus scheme, the pattern of gaugino masses in mixed modulus-K\"ahler mediation is more compressed. The unification of the gaugino masses at the GUT scale is modified due to the non-universal contribution from $b^\prime_a$. Mirage unification of the gaugino (scalar) masses is affected as well.\footnote{Strictly speaking, it is not possible to define a mirage scale unambiguously in this context since the splitting and RG evolution of the gaugino masses are due to different coefficients.}

We have to keep in mind that $\phi=\mathcal{O}(1)$ could be problematic since model dependent contributions like (string) threshold corrections sensitively depend on the value of $\phi$ \cite{Choi:2007ka} and would not allow for an unambiguous identification. This would even stronger affect the scalar masses. Thus in order to have a controllable scheme we have to keep the value of $\phi$ small (i.\,e.\ find models that admit $\phi\ll1$). 

Eventually, the general mixed mediation scenario is given in Eq.\,(\ref{generalpattern}). Certainly it might be difficult 
to identify the scheme if all contributions are present. However, we can use the above argument and consider small $\phi$. This would then lead to a deviation from the pure mirage mediation scenario. The three gaugino mass curves cut out a ``mirage triangle'' (see Fig.\,\ref{fig:kahler}) around the mirage scale Eq.\,(\ref{miragescale}).

In conclusion we see that in mixed schemes with several parameters it will be difficult (if not impossible) to obtain further evidence for grand unification from the measurement of soft parameters at the LHC. We have to hope for simple (controllable) schemes like gravity/modulus mediation, anomaly mediation and a mixture thereof (mirage mediation).

\section{Constraints from phenomenological requirements}

\begin{figure}
\captionsetup[figure]{labelfont={footnotesize,bf},textfont=footnotesize,labelsep=mysep,labelformat=mypiccap,format=default,justification=RaggedRight,width=15cm,indent=5pt}
\begin{center}
\begin{minipage}{0.74\linewidth}
\includegraphics[width=\linewidth]{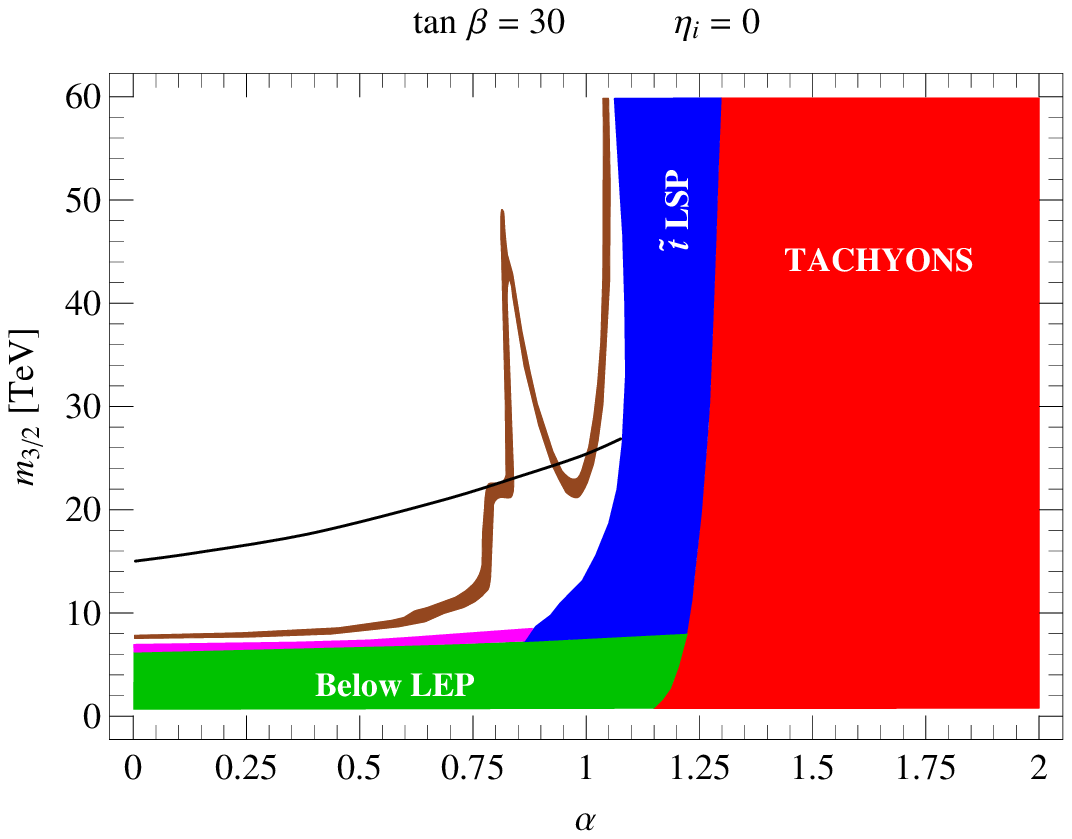}
\caption{Constraints on the $\lbrace\alpha,m_{3/2}\rbrace$ parameter space in pure mirage mediation. In the brown strip the thermal neutralino abundance satisfies the WMAP bounds. Below the strip the thermal abundance is too low whereas above it is too large. In the pink area (above the LEP region) the stau is the LSP. The area below the black curve is excluded by the $b\rightarrow s\gamma$ constraint. The public codes \texttt{SOFTSUSY}\cite{Allanach:2001kg} and \texttt{micrOMEGAs} \cite{Belanger:2001fz} have been used.}
\label{fig:space1}
\end{minipage}
\end{center}
\begin{center}
\begin{minipage}{0.74\linewidth}
\includegraphics[width=\linewidth]{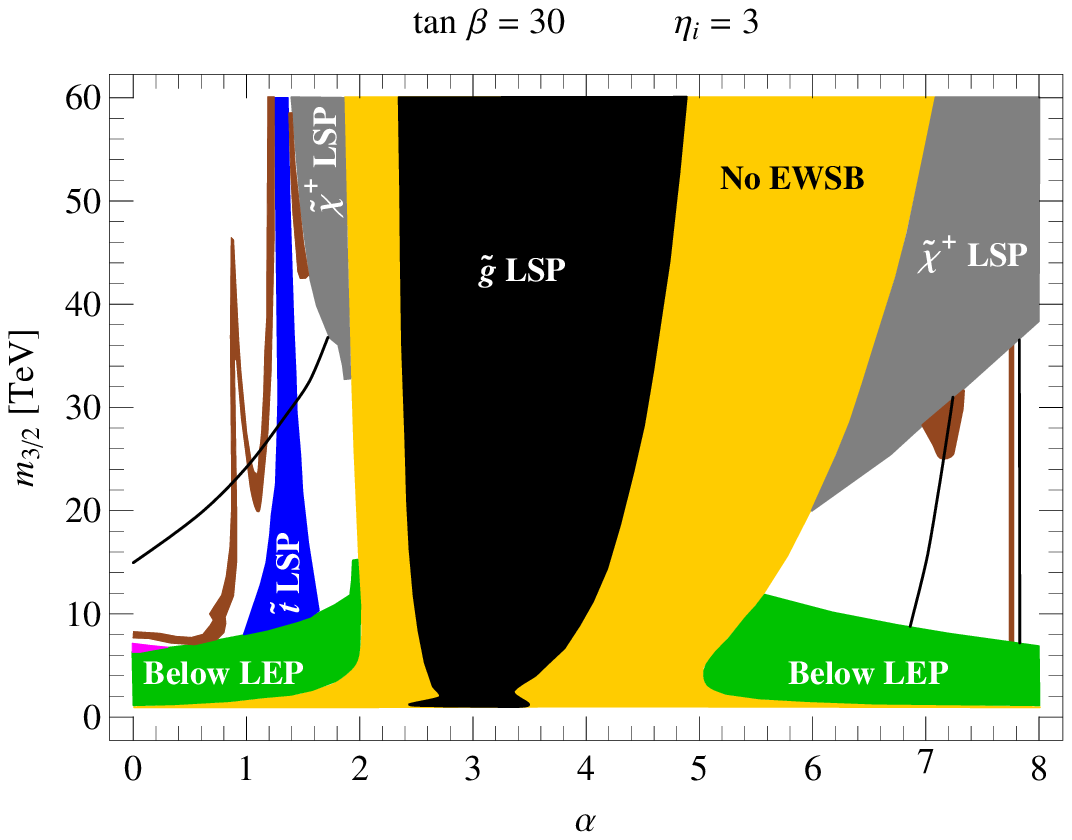}
\caption{Same as Fig.\,\ref{fig:space1}. The area between the two black curves on the right side is excluded by the \mbox{$b\rightarrow s\gamma$} constraint.}
\label{fig:space2}
\end{minipage}
\end{center}
\end{figure}

But even these simple schemes have to fulfill further phenomenological conditions, like the absence of tachyons, satisfactory electroweak symmetry breakdown, consistency with LEP data and cosmological constraints. In the schemes under consideration this will correspond to the search for allowed ranges of $\alpha$. We shall not discuss these questions here in full detail as this will lead to quite some model dependence \cite{Endo:2005uy,Choi:2005uz,Falkowski:2005ck,Baer:2006id,Baer:2006tb,Choi:2006im}. Instead we shall just consider the simplest cases.

The first nontrivial realization of mirage mediation has been identified within Type IIB string theory and a specific uplifting sector \cite{Kachru:2003aw}. It is characterized essentially by two parameters: $\alpha$ and $m_{3/2}$. This is even more constrained than the simple gravity mediation scheme: in particular we have $m_0=m_{1/2}$. The value of $\alpha$ depends on the uplifting mechanism. The original suggestion gave $\alpha=1$. In Fig.\,\ref{fig:space1} we show the allowed range of parameter space in the $\lbrace \alpha, m_{3/2}\rbrace $ plane. It is obvious that in this case large values of $\alpha$ are forbidden since in that region we are close to the scheme of anomaly mediation which suffers from tachyonic sleptons. We see that $\alpha$ is confined to rather small values to avoid these tachyons as well as the region with a stop-LSP. Low values of $m_{3/2}$ are forbidden from LEP data \cite{Yao:2006px} and the constraints from $b \rightarrow s\gamma$ \cite{Tajima:2001qp}. The brown strip indicates the region where the thermal relic abundance of the LSP-neutralino is consistent with the WMAP result \cite{Spergel:2006hy}. We see that a large part of the parameter space is ruled out, but it is still interesting to see that this simplest model with $\alpha=1$ survives all the constraints.

To get to larger values of $\alpha$ one has to consider additional contributions to the scalar masses to remove the tachyonic sleptons. We consider here the situation of $F$-term uplifting as described in \cite{Lebedev:2006qq}. The scalar masses are then given by:
\begin{align}
\nonumber m^2_i &= m^2_{1/2}\Bigg[ 1 - \frac{\eta^2_i}{(16\pi^2)^2} + \alpha^2\,\frac{\log^2(M_{\rm P}/m_{3/2})}{(16\pi^2)^2}\,\eta^2_i \\
&\phantom{=m^2_{1/2}\Bigg[ 1}-\alpha^2\,\frac{\log^2(M_{\rm P}/m_{3/2})}{(16\pi^2)^2}\,\dot{\gamma}_i + 2\hs\alpha\,\frac{\log(M_{\rm P}/m_{3/2})}{16\pi^2}\,\Psi_i \Bigg]\,,
\end{align}
where $\eta_i$ denotes the additional contribution from the uplifting sector to the scalar masses and the quantity $\Psi_i$ is related to the moduli dependence of the gauge couplings (see \cite{Choi:2004sx,Choi:2005ge} for details).

The allowed range of parameter space for $\eta_i=3$ is shown in Fig.\,\ref{fig:space2}. Apart from the window at small $\alpha$ we now have also an allowed range for larger $\alpha>6$. In between there are problems with electroweak symmetry breakdown and also a region with a gluino LSP. Consistency with WMAP data requires an upper limit on $\alpha$. Many more specific models can be considered that allow wider ranges for $\alpha$ and $m_{3/2}$. We shall not discuss them here in detail.

\section{Outlook}

Upcoming experiments at the LHC might test the paradigm of Grand Unification (GUT) in particle physics. In principle
we could be confronted with three different situations:
\begin{enumerate}
\item[1)] GUTs are ruled out by the data
\item[2)] Experimental results support GUTs
\item[3)] One can still build models that eventually can be made consistent with GUTs.
\end{enumerate}

Our experience with ingenuity in particle physics model building makes us confident that the third possibility can practically not be ruled out. A posteriori we might set up models with specific boundary conditions and new particles at intermediate scales that lead to unification of gauge couplings at some scale.

Therefore it is important to define some a priori criteria that would support either 1) or 2). In this paper we have 
analyzed those that could serve as arguments for 2).\footnote{It would be of interest to analyze the options for 1) as well.} We have selected gaugino masses as the most promising tools for this analysis. They are quite often directly
related to the gauge coupling constants and are less model dependent than other soft parameters. A meaningful analysis
also requires the consideration of simple boundary conditions, and simple mechanisms of the mediation of supersymmetry
breakdown form hidden to observable sector.

Given all this, mass ratios of gauginos might provide ``smoking gun'' signals in favor of GUTs. The simplest ones are those obtained in gravity mediation with universal boundary conditions:
\begin{align}
M_1:M_2:M_3 &\simeq 1:2:6\,.
\end{align}
Observation of this ratio will provide a strong support for grand unification. Another ratio which is quite special
reads
\begin{align}
M_1:M_2:M_3 &\simeq 3.3:1:9\,,
\end{align}
a clear signal for anomaly mediation. More complicated schemes require information from soft parameters other than the
gaugino masses alone and the identification of the scheme: support for grand unification becomes more and more difficult. There is hope that a mixture of gravity and anomaly mediation, mirage mediation, can be identified via LHC data. The mixing parameter $\alpha$ might be determined through the knowledge of gaugino masses and soft scalar masses. Beyond this, unambiguous signals for grand unification, if possible at all, would require precision data beyond that available at the LHC. New suggestions for a priori criteria in favor of GUTs would be welcome. Alternatively one might try to formulate solid criteria against the idea of grand unification before LHC comes into operation.

\vspace{0.6cm}
\subsection*{Acknowledgements}

This work was partially supported by the European Union 6th framework program MRTN-CT-2006-035863 ``UniverseNet'' and 
SFB-Transregio 33 ``The Dark Universe'' by Deutsche Forschungsgemeinschaft (DFG). 
\vspace{0.6cm}

\addcontentsline{toc}{section}{{Bibliography}}

\providecommand{\href}[2]{#2}\begingroup\raggedright
\endgroup

\end{document}